# Simpsonian "Evolution by Jumps" in an Adaptive Radiation of *Anolis* Lizards

Jonathan M. Eastman[1], Daniel Wegmann[2], Christoph Leuenberger[3], and Luke J. Harmon[1*]

[1] Department of Biological Sciences & Institute for Bioinformatics and Evolutionary Studies (IBEST), University of Idaho Moscow, ID 83843-3051

[2] Department of Biology, Fribourg University, Fribourg, Switzerland

[3] Department of Mathematics, Fribourg University, Fribourg, Switzerland

* corresponding author: lukeh@uidaho.edu

**ABSTRACT**

**In his highly influential view of evolution, G. G. Simpson hypothesized that clades of species evolve in adaptive zones, defined as collections of niches occupied by species with similar traits and patterns of habitat use. Simpson hypothesized that species enter new adaptive zones in one of three ways: extinction of competitor species, dispersal to a new geographic region, or the evolution of a key trait that allows species to exploit resources in a new way. However, direct tests of Simpson's hypotheses for the entry into new adaptive zones remain elusive. Here we evaluate the fit of a Simpsonian model of "jumps" between adaptive zones to phylogenetic comparative data. We use a novel statistical approach to show that anoles, a well-studied adaptive radiation of Caribbean lizards, have evolved by a series of evolutionary jumps in trait evolution. Furthermore, as Simpson predicted, trait axes strongly tied to habitat specialization show jumps that correspond with the evolution of key traits and/or dispersal between islands in the Greater Antilles. We conclude that jumps are commonly associated with major adaptive shifts in the evolutionary radiation of anoles.**

**INTRODUCTION**

G. G. Simpson (1944) postulated that lineages typically evolve and speciate within adaptive zones, defined as collections of niches occupied by species with similar traits and patterns of habitat use. Simpson proposed that lineages occasionally transition from one adaptive zone to another, and that these transitions occur during brief periods of rapid evolution, which he called "quantum evolution" (Simpson 1953). Quantum evolution was intended to resolve what Simpson regarded as the most important and outstanding theoretical problem of the fossil record: the geologically sudden appearance of new forms. According to Simpson, these abrupt changes reflect lineages' rapid transitions into new adaptive zones.

In Simpson's framework, transitions into new adaptive zones can happen for one of three reasons. First, mass extinction of species can create an ecological opportunity that is then exploited by a new evolutionary lineage. Second, dispersal can create a geographic opportunity when species arrive in a new geographic area that is devoid of potential competitors (e.g. on Madagascar, where several lineages have diversified rapidly; see Yoder and Nowak 2006). Third, species can evolve a phenotypic novelty (i.e., key innovation) that allows the lineage access to new resources that were previously unavailable. A classic example of this is the evolution of flight, which is hypothesized to have spurred subsequent evolutionary radiation in mammals (bats; Jones et al. 2005), reptiles (birds; Padian and Chiappe 1998), and insects (Hennig 1981).

There have been few direct tests of Simpsonian jumps between adaptive zones using empirical data (but see Uyeda et al. 2011). Here, we analyze "evolution by jumps" in the adaptive radiation of anoles, lizards that have adaptively radiated in the Caribbean and South America (reviewed in Losos 2009). Following previous work, we focus on anoles on the four islands of the Greater Antilles. These anoles provide a unique opportunity for testing Simpson's theory of adaptive zones for two reasons. First, there have been repeated dispersal events among islands in the Greater Antilles (Losos et al. 1998; Losos 2009; Nicholson et al. 2005). These dispersal events potentially represent geographic opportunities, where anole lineages reach a new island and are no longer sympatric with the former set of competitors (Mahler et al. 2010). Second, most anole species can be classified into ecomorphs, habitat specialists that have evolved repeatedly on the four islands of the Greater Antilles (Losos et al. 1998). Transitions between ecomorph categories represent the evolution of key characters in anole lineages that allow them to invade novel habitats (Losos et al. 1998; Harmon et al. 2005; Losos 2009). Thus, anoles have repeatedly meet two of Simpson's three conditions expected to be associated with evolutionary jumps. We know of no obvious extinction events that might have affected anole evolution, so we cannot address this third potential factor for promoting evolutionary jumps in anoles. Importantly, both ecomorph origins and transitions among islands are replicated in the phylogeny of anoles, but are still rare enough that we can estimate the position of transitions on the phylogenetic tree with some confidence (Schluter 1995; Huelsenbeck et al. 2003; see also Appendix A).

We developed and applied a novel method to address two fundamental questions about evolution by jumps in anoles. We first determine whether jump models fit better than alternative uniform models of trait evolution. Secondly, we ask whether jumps correspond with either of the two factors postulated by Simpson: evolution of key characters and/or geographic dispersal.

To address these questions, we analyze data on anole body size and shape in conjunction with a time-calibrated phylogenetic tree of these lizards (Mahler et al. 2010). We focus on snout-vent length (SVL), a standard phenotypic measurement of body size for lizards, along with major axes of body shape variation (see below). Together these traits are broadly correlated with habitat partitioning in Greater Antillean anoles and represent the primary axes of ecologically driven evolutionary divergence in these lizards (Schoener 1970; Beuttell and Losos 1999; Losos 2009).

**MATERIAL and METHODS**

*Morphometric data and phylogeny*

We use a morphometric dataset and phylogeny estimates for Greater Antillean anoles to test Simpson's model of quantum evolution in a radiation of lizards. The phylogeny we use comprises roughly 1500 aligned sites in length, involving seven markers within the mitochondrial genome (ND2, trnW, trnA, trnN, trtC, trnY, and the beginning of COX1; Mahler et al. 2010). From these data, Mahler et al. (2010) sample 898 trees from the posterior density using an uncorrelated relaxed-clock model in BEAST (Drummond and Rambaut 2007). For our analyses, we rescale a random sample of 100 trees from the posterior density of Mahler et al. (2010) to an arbitrary root height of 1. Morphometric data include 22 characters measuring bone lengths, head shape, body shape, body size, and characteristics of the lamellae (toepads). The tree and morphometric datasets includes roughly 84 percent of the known diversity in the Greater Antilles (sampling 100 of 119 total known species from Cuba, Jamaica, Hispaniola, and Puerto Rico; Losos 2009). We assess fit of jump-diffusion, diffusion, and relaxed-diffusion models of evolution for body size (natural-log transformed snout-to-vent length, hereafter 'SVL'), along with principal-component axes of body shape. Following Mahler et al. (2010), we use phylogenetic principal component analysis ('pPCA'; see Revell 2009) to resolve major axes of variation in a body shape dataset, assuming the correlation structure among species to be in proportion to their shared ancestry, following a multivariate Brownian motion model of evolution. We first find the residuals from a regression of the measured characters against body size using phylogenetic generalized least squares, then apply a pPCA to these body shape residuals, retaining the first five axes for further analysis (Table 1).

*Estimating ecological opportunity*

We use a model selection approach to test Simpson's (1944, 1953) basic claim that entry into a novel adaptive zone is concomitant with episodic bursts of rapid ('quantum') evolution. We construct a set of candidate models to compare amongst: these models are with or without jumps, and focal set of jump models restrict jumps to certain branches in the tree where major ecomorphological transitions and successful inter-island dispersal events have taken place. To reconstruct these events with respect to phylogeny, we use stochastic mapping (Huelsenbeck et al. 2003) as implemented in the R package PHYTOOLS (v0.1.7; Revell 2012). PHYTOOLS uses the maximum-likelihood estimate of the transition matrix to stochastically generate character states and character state transitions at nodes and along branches of the tree. We reconstruct transitions among ecomorphs (representing transitions in key characters) and islands (representing changes in geography) using this method. We limit the number of model parameters by assuming an 'equal-rates' matrix between transition types for transitions between ecomorph classes and islands. For each draw from our posterior subsample of 100 trees, we generate a stochastic mapping for ecomorphology and for geography. Each stochastic mapping defines a set of branches where transitions in key characters or geography occurred. According to our Simpsonian hypothesis of entry into novel adaptive zones, we hypothesize that these transitions should correspond with jumps in morphospace, which we model using a jump-diffusion process.

*Jump-diffusion model description*

We develop a novel statistical approach based on Levy flight models (Viswanathan et al. 1996; Landis et al. 2013) and classic economic theory (Merton 1976) to detect evolutionary jumps using phylogenetic comparative data. Our model supposes that evolutionary change follows a 'random walk,' modeled by Brownian motion governed by a walk-variance parameter, $\sigma^2_W$. Rarely, lineages experience instantaneous jumps, where the amount of evolutionary change is drawn from a normal distribution with mean 0 and a jump variance parameter, $\sigma^2_J$. Under the 'free' version of this model, we allow jumps to occur along any branch in the phylogenetic tree relating species; alternatively, we restrict jumps to occur only on those branches predicted to be associated with Simpsonian transitions into new adaptive zones (see below).

We calculate the likelihood of a jump model given a phylogenetic tree and data for a continuous character, measured for all species on the tips of that tree. Under this model, species traits follow a multivariate normal distribution (see Appendix A for full derivation), with:

$$E[y_i] = \mu$$
$$cov[y_i, y_j] = \sigma^2_{ij} = \tau_{ij}\sigma^2_W + n_{ij}\sigma^2_J$$

Here, $y_i$ represents the trait value for species $i$, $\mu$ the ancestral trait value of the group, $cov(y_i, y_j)$ the covariance between species $i$ and $j$, $\tau_{ij}$ the shared branch length between species $i$ and $j$, $n_{ij}$ the number of jumps that occur on the shared branches of species $i$ and $j$, and $\sigma^2_W$ and $\sigma^2_J$ the walk and jump variance, respectively (Cavalli-Sforza and Edwards 1967; Felsenstein 1973).

*Model fitting*

We define three Simpsonian jump-diffusion models: a *geographical* model, where jumps are expected only along branches where colonization of a new geographic region (i.e., an island); an *ecomorphological* model, where jumps are restricted to branches along which transitions between ecomorphological classes occur (treating each unique anole as its own ecomorph); and an *ecomorphological-geographical* model, where jumps are expected if lineages transitioned either in ecomorphology or in geographic setting. We identify branches where jumps can occur using stochastic character mapping (as above). We then compare the fit of these three 'restricted' Simpsonian models with three alternatives: a single-rate diffusion model (*Brownian motion*; Felsenstein 1985), a relaxed diffusion model (*relaxed Brownian-motion*; O'Meara et al. 2006; Revell et al. 2012; Eastman et al. 2011), and a *free* jump-diffusion model. The free model is a jump-diffusion model in which there is no limitation in the topological position(s) of inferred jumps. In contrast, the *relaxed Brownian-motion* model supposes that jumps are absent but that diffusion rates may vary across the tree (see Eastman et al. 2011).

We use a Markov chain Monte Carlo (MCMC) algorithm to sample from the posterior distribution of the parameters of each Bayesian model. For jump models, jump locations are considered latent variables, and their number and placement on the tree is adjusted as part of the MCMC algorithm (see Appendix B for further detail).

*Model selection*

We use Bayes factors to compare competing models of trait evolution in anoles. We calculate Bayes factors from estimated marginal likelihoods using a steppingstone approach (Xie et al. 2011). Several other methods exist to estimate the marginal likelihood of a model. The harmonic mean estimator (HME) of the marginal likelihood (Newton and Raftery 1994) has been a commonly employed and simple method for model selection in Bayesian inference. While popular, the practice of using the HME is deeply flawed given that the estimator is often biased and may suffer from infinite variance (e.g., see Fan et al. 2011). Recently, the novel approach of thermodynamic integration has been proposed by Lartillot and Philippe (2006) and by Friel and Pettitt (2008) as a more robust approach. This method, which uses power transformations of the model probabilities (i.e., 'power posteriors'), was extended by Xie et al. (2011) for steppingstone analyses. In the latter method, a partitioned estimate of the marginal likelihood is obtained by computing a series of ratios of normalizing constants for pairs of power posteriors spanning from the prior to the posterior density. The steppingstone approach makes use of the arithmetic mean estimator of the marginal likelihood, which is most stable when the importance distribution is slightly more diffuse than the target distribution. This outcome is exploited by the steppingstone approach (Xie et al. 2011). When model selection is a key component of the analysis, this more direct (yet more computationally costly) steppingstone approach is recommended as a marginal model-likelihood estimator (see also Fan et al. 2011).

We apply the steppingstone estimator in model comparison for our anole dataset. Our model-selection inferences are integrated over phylogenetic uncertainty as well as uncertainty in the precise topological locations of transitions between ecomorph categories and inter-island colonization events within the Greater Antilles. We use 5 $\beta$-power posteriors with $\beta$ values equally spaced between 0 to 1, forming a progression of distributions that fully range from the prior to the posterior. We use Markov chains for each power posterior that are $10^6$ generations in length, the first half of which allow the chain to burn-in. Thereafter, we retain a sample at intervals of 200 generations. Chains are evaluated for stationarity using the Heidelberger and Welch (1983) test and visual inspection using CODA.

Estimation of the log marginal model-likelihoods from each set of power posteriors follows Xie et al. (2011). Using Bayes factors, we compare support for several competing models of trait evolution in a pairwise manner. We interpret model preferences following recommendations of Kass and Raftery (1995), using twice the difference in estimated marginal log-likelihood for two competing models (i.e., $2\log_e B$, where $B$ is the Bayes factor or marginal-likelihood ratio of the two models). To summarize model preferences, we calculate model weights by determining the relative marginal likelihood for each Bayesian model across the distribution of 100 trees (Table 1).

*Implementation and testing*

Our methods are implemented in the R package GEIGER (Harmon et al. 2008; Eastman et al. 2011). Simulations confirm that this method has satisfactory statistical properties, including high power to detect jumps when they occur and a low false-positive rate (see Appendix B). A full description of our data analyses, including prior probability and proposal distributions used in MCMC sampling, can be found in the Supplementary Information (sections 1-2).

**RESULTS**

We find strong evidence for jump-diffusion models of evolution for three axes of morphometric variation in anoles, two of which provided strong evidence in favor of Simpson's hypotheses of transitions among adaptive zones (Table 2). Figures summarize inferences of jump locations for the preferred evolutionary models of body size (Fig. 1) and the first and third principal component axes (Figs. 2-3), datasets for which preference for a single model is substantial (Table 2). We compile posterior samples of jump locations across the distribution of trees and plot averages of posterior probabilities amongst these samples. Note that each unique branch in the posterior sample of trees may not be present in the maximum clade credibility tree plotted in Figs. 1-3. Nonetheless, the most highly supported jumps appear on branches present in the summary tree.

Simpsonian models of phenotypic evolution are by far the most commonly preferred models in anoles (Table 2). For body size, almost the entire model weight for the phenotypic axis of body size is apportioned between two Simpsonian jump models (Table 2), either allowing jumps only at ecomorphological transitions or at ecomorphological transitions as well as inter-island dispersal events. Moderately strong model preferences for these two models are similarly observed for first and third principal component axes (PC1: relative lengths of the extremities and long bones; PC3: contrasts tail length and width of the lamellae; Table 1). Other trait axes are ambiguous with respect to model preference (Table 2).

**DISCUSSION**

Our analyses support two main conclusions. First, evolutionary change in anoles is not well described by a uniform random walk. A better description of anole evolution combines a uniform component of change that is punctuated by rapid jumps in trait values. These jumps have been relatively common in the history of anoles, affecting three of six measured morphological axes. Second, for three trait axes, jumps in trait space correspond to ecological transitions to novel ecomorphs and perhaps also island dispersal events. The evolution of these characters is consistent with Simpson's description of evolutionary jumps associated with the entry into new adaptive zones.

Our results also provide support for Simpson's model of adaptive zones, finding substantial support for jumps in trait evolution that follow Simpson's quantum evolutionary hypothesis as lineages enter new adaptive zones. These results also show that the distinction between "gradual" and "punctuated" models of evolution is a false dichotomy; instead, evolution has a gradual component that may be frequently punctuated by periods of rapid change (Levinton 1988).

The strongest support for Simpsonian evolution by jumps in anoles was for body size. Body size is well regarded as a fundamental trait of an organism, inextricably tied to the modes and intensities of interspecies interaction as well as ecological properties of a species (Schoener 1969, 1974; Werner and Gilliam 1984; Woodward et al. 2005). In anoles, body size is highly correlated with many aspects of functional performance and habitat usage (Losos 2009). Support for the ecomorphological model for Greater Antillean body sizes consistent with evolution of anoles elsewhere in the Carribean (Table 2; Fig. 1). Smaller islands in the Lesser Antilles with only two anole species seem to diverge along a primary axis of body size (Schoener 1970; Roughgarden 1995), yet many of these anoles cannot be unequivocally assigned to the six recognized ecomorph classes found in the Greater Antilles. Taken together, it seems probable that body size is a primary axis of trait divergence consonant with ecological differentiation.

Toe pads have likely been a central feature of the anole radiation (Losos 2009; Pinto et al. 2008). Highly variable in number and width, features of the lamellae are well correlated with arboreality (Collette 1961). Generally well developed in the Greater Antillean anoles and highly evolutionarily labile, toe-pad evolution is thought to undergird adaptive radiation in these communities (Losos 2009), as in other lineages of lizard (geckos and skinks: Irschick et al. 2006; Williams and Peterson 1982). Our results lend support to the importance of toe pads in the anole adaptive radiation. We find good support for Simpsonian models of quantum evolution for the morphometric axes of variation strongly influenced by characteristics of the lamellae, PC1 and PC3 (Tables 1-2; Figs. 2-3).

A poorly understood phenomenon in the Greater Antilles is the appearance of several lineages of 'unique' anoles alongside strong patterns of convergence in communities of anoles in these same four islands. Many unique anoles are syntopic with some of the ecomorphs, others are riparian, still others are saxicalous, and one is semi-aquatic. It is remarkable that nearly all the unique anoles occur on the two largest islands of the Greater Antilles, Hispaniola and Cuba (Losos 2009). A single unique anole (*A. reconditus*) occurs in Jamaica, and Puerto Rico has none. One further peculiarity of these anole communities appears to be the close relatedness of unique anoles: these species are more closely related on average than would be expected from random distribution across the phylogeny (p < 0.05).

It has been speculated that the niches occupied by unique anoles are minor 'peaks' in the adaptive landscape of anoles, exploited only after the six ecomorph classes are fully occupied (Losos 2009). Yet the apparent long durations of several unique anole lineages would make this seem improbable (Losos 2009). It could be that the breadth of an adaptive zone (i.e., in terms of the number of species that can be supported) might not be correlated with its accessibility. That is, adaptive zones that support few species (e.g. possibly a single 'unique' anole) may also happen to be one of the more accessible zones. For the body size dataset, support for the ecomorphological model appears largely driven by well supported jumps inferred along branches leading to unique anoles (Fig. 1). At least for this axis of morphological variation, pulses of evolution in the unique lineages appear similar (if not greater) in frequency than for the six recognized ecomorph classes (Fig. 1).

Although we model evolutionary jumps as instantaneous, we want to be clear that we are not invoking actual instantaneous evolutionary change (e.g. "hopeful monsters"; Goldschmidt 1940; Charlesworth et al. 1982). Typical microevolutionary processes of selection and drift can cause change that would appear to be instantaneous when viewed over the timescale of anole evolution, measured in tens of millions of years[8]. Our model is also distinct from punctuated equilibrium, which requires evolutionary jumps to occur only at speciation events (Eldredge and Gould 1972). The punctuated changes in our model occur along branches in the tree and are not necessarily associated with speciation events. In fact, two lines of evidence argue against punctuated equilibrium in anoles: first, most speciation events in our tree are not associated with jumps in any of our six characters; and second, we know from detailed microevolutionary studies that anole body size and shape can evolve rapidly in response to selection even in the absence of speciation (Losos et al. 2006).

In summary, we find strong evidence for Simpsonian 'evolution by jumps' in the evolution of anoles in the Greater Antilles. We also provide statistical support for Simpson's model of transitions among adaptive zones associated with the evolution of key characters and/or geographic opportunity. We suggest that future work should follow Simpson's lead and focus on the factors that promote pulses of evolutionary change.

**ACKNOWLEDGEMENTS**

We are indebted to Jonathan Losos, Matt Pennell, Graham Slater, Joseph Brown, Michael Alfaro, and Luke Mahler for providing invaluable support throughout development of the project. Special thanks are due to Rob Lyon, without whom many of the analyses would have been impractical. Funding for the project was provided by NSF DEB 0919499 and NSF DEB 1208912 (LJH). JME and LJH appreciated support through NESCent (NSF EF-0905606). The funders had no role in study design, data collection and analysis, decision to publish, or preparation of the manuscript.

**COMPETING INTEREST**

The authors declare that no competing interests exist.

**FIGURES**

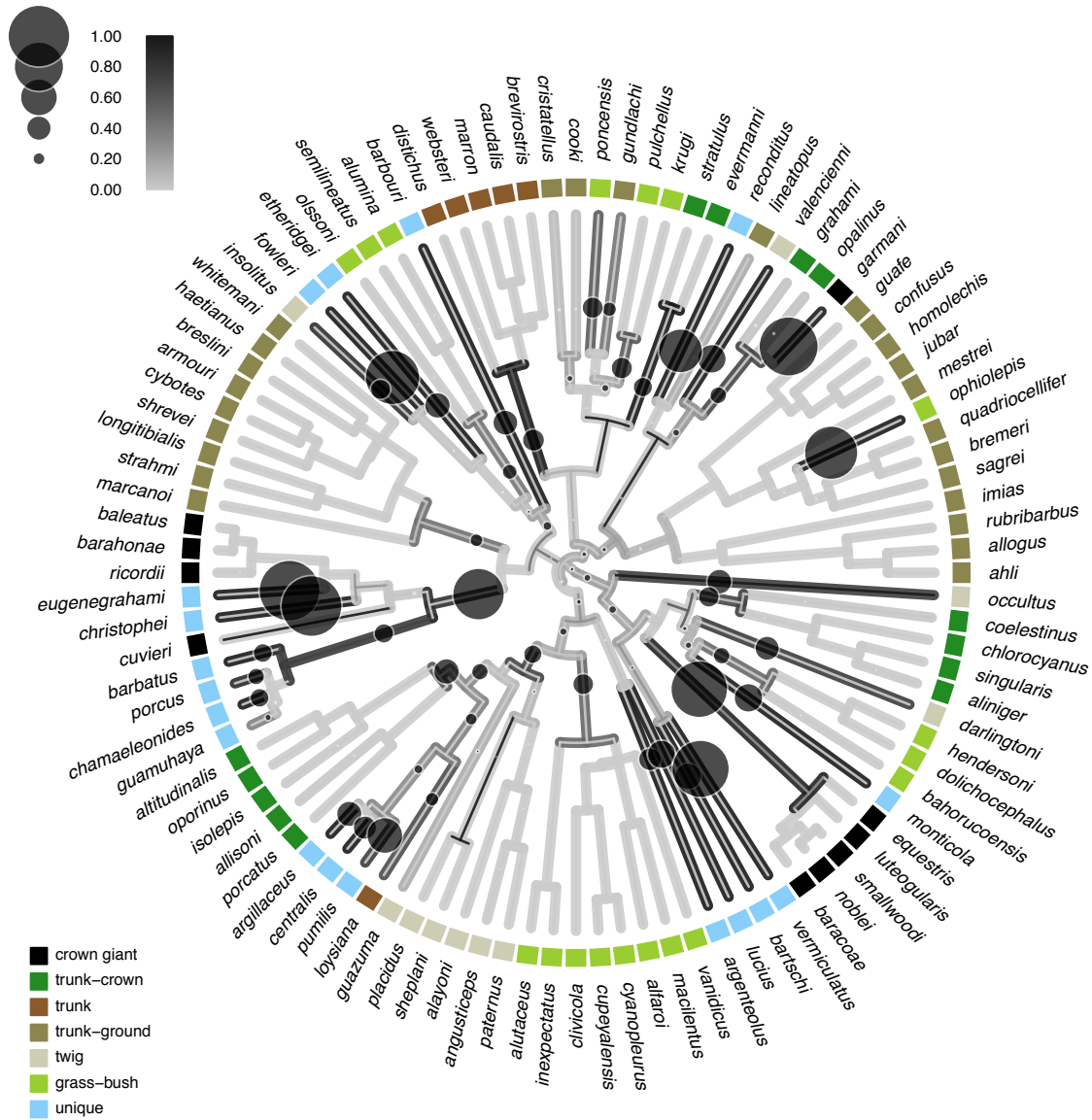

Figure 1. *Anolis* summary tree showing inferred jumps in body-size evolution. Jump inferences are summarized for the preferred *ecomorphological* model (model weight = 0.68; Table 2), permitting jumps only where transitions in ecomorphology have been estimated by stochastic mapping. Estimates for jump and transition probabilities are marginalized across 100 trees and are summarized here on a maximum clade-credibility tree of Mahler et al. (2010). Circles are sized in proportion to the marginal posterior probability of a jump along that branch, following the uppermost legend. Line shadings of branches reflect the average conditional probabilities of transitions in ecomorphology (outer) and geography (inner) across the set of 100 trees; shadings follow the uppermost legend. Median jump effect size (jump variances scaled by the random-walk variance) is estimated to be 25.45. Median estimate of the proportion of all branches experiencing jumps is 0.108. Results for PCI and PCIII appear in the Supplementary Information (section 5).

Figure 2. Anole tree with jumps in morphological space according to the preferred *ecomorphological-geographical* model for PCI (model weight = 0.49; Table 2). Median intensity of the jump process (α) is estimated to be 7.83; the median frequency of branches experiencing jumps is estimated to be 0.145. Other details follow Figure 1.

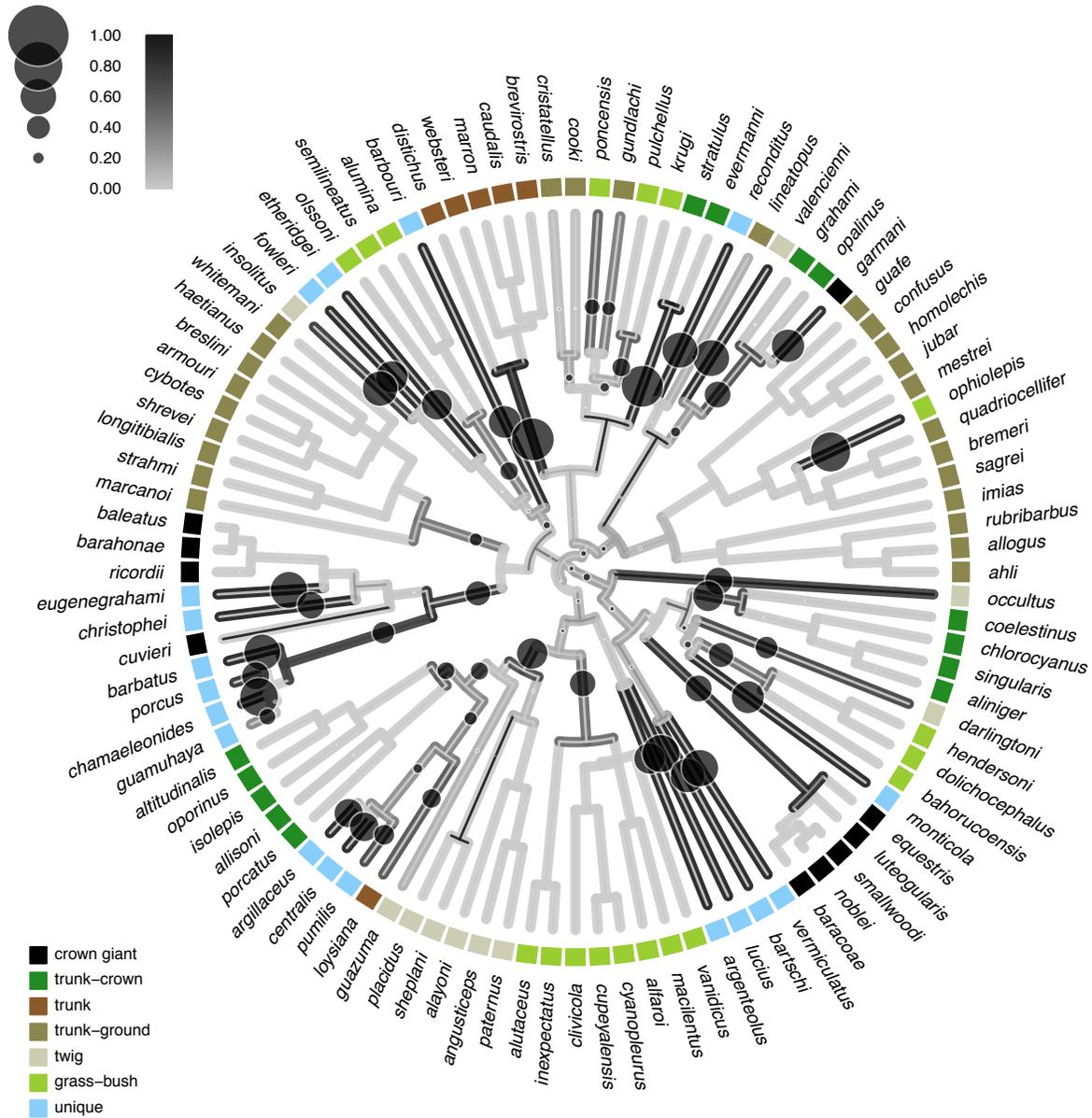

Figure 3. Anole tree with jumps in morphological space according to the preferred *ecomorphological* model for PCIII (model weight = 0.39; Table 2). Median intensity of the jump process (α) is estimated to be 16.26 and with a median estimated proportion of branches experiencing jumps of 0.135. Other details follow Figure 1.

# TABLES

Table 1. Loadings of the original phenotypic measures on the size-corrected, principal component axes used in this study. Proportions of total variation in the size-corrected data are 0.37, 0.21, 0.12, 0.10, and 0.04 for the respective principal component axes. Measurements pertaining to bones of the hand (or foot) are taken from the fourth finger (or toe). Further details of each measurement can be found in Mahler et al. (2010).

|  | PCI | PCII | PCIII | PCIV | PCV |
|---|---|---|---|---|---|
| head length | 0.18 | -0.92 | -0.14 | 0.04 | -0.13 |
| head width | -0.61 | -0.58 | 0.18 | -0.34 | 0.13 |
| head height | -0.25 | -0.62 | 0.15 | -0.49 | 0.17 |
| jaw length | 0.10 | -0.91 | -0.26 | 0.14 | -0.04 |
| jaw opening in-lever | 0.17 | -0.92 | -0.24 | 0.15 | -0.10 |
| jaw closing in-lever | 0.22 | -0.83 | -0.30 | 0.25 | -0.19 |
| femur length | -0.84 | 0.09 | -0.35 | 0.05 | -0.11 |
| tibia length | -0.85 | 0.12 | -0.39 | 0.13 | -0.09 |
| metatarsal length | -0.85 | 0.10 | -0.45 | 0.13 | -0.02 |
| toe length | -0.83 | 0.02 | -0.38 | 0.19 | 0.05 |
| pelvis height | -0.25 | -0.06 | 0.20 | 0.84 | 0.24 |
| pelvis width | -0.87 | 0.03 | 0.17 | -0.15 | -0.29 |
| humerus length | -0.78 | 0.07 | 0.02 | -0.01 | -0.40 |
| radius length | -0.89 | -0.01 | -0.05 | 0.02 | -0.07 |
| metacarpal length | -0.53 | -0.26 | 0.68 | 0.11 | 0.01 |
| finger length | -0.28 | -0.18 | 0.51 | 0.67 | 0.09 |
| foot lamellae | -0.53 | -0.13 | 0.64 | 0.17 | 0.11 |
| hand lamellae | -0.82 | 0.07 | 0.19 | -0.19 | -0.11 |
| toe lamella width | -0.53 | -0.30 | -0.23 | -0.37 | 0.51 |
| finger lamella width | -0.61 | -0.28 | 0.40 | -0.23 | 0.12 |
| tail length | -0.42 | 0.20 | -0.64 | 0.18 | 0.40 |

Table 2. Model weights for each morphological axis derived using the steppingstone method and marginalized across tree space.

|  | SVL | PCI | PCII | PCIII | PCIV | PCV |
|---|---|---|---|---|---|---|
| Free | 0.00 | 0.25 | 0.17 | 0.09 | 0.16 | 0.18 |
| Ecomorphological-Geographical | 0.31 | 0.49 | 0.17 | 0.27 | 0.18 | 0.16 |
| Ecomorphological | 0.68 | 0.15 | 0.17 | 0.39 | 0.19 | 0.16 |
| Geographical | 0.00 | 0.05 | 0.16 | 0.08 | 0.15 | 0.15 |
| Relaxed Brownian Motion | 0.00 | 0.02 | 0.16 | 0.08 | 0.15 | 0.19 |
| Brownian Motion | 0.00 | 0.02 | 0.16 | 0.09 | 0.17 | 0.17 |

**APPENDIX A**

**Model description**

Here we develop a general model of continuous trait evolution inspired by Simpson's (1944, 1953) theory of quantum evolution. While Simpson's quantum evolution has resisted formulation as an explicit model, a recent broad study of trait divergences is strongly suggestive of rare pulses in morphological evolution (Uyeda et al. 2011). Below, we introduce a model of trait evolution that superimposes an instantaneous 'jump' process on a continuous-time diffusion process.

We consider the distribution of traits among species related by a time-scaled phylogenetic tree in our 'jump' model. Consider the observed phenotypic values (**y**) for $N$ species:

$$\mathbf{y} = (y_1, ..., y_N) \qquad (A1)$$

Our model predicts the elements of **y** will follow a multivariate normal distribution with mean $\mu$ and variance **S**. Here, µ is the root state at which the process begins and **S** is the variance-covariance matrix of phenotypic values among species. A fully resolved and rooted tree ($\tau$) will have $K = 2N$-2 total branches (excluding the root). In this model, µ is a column vector repeating the ancestral phenotypic value $N$ times, and **S** is the $N$-by-$N$ dimensional variance-covariance matrix. Each element within **S** = [$s_{ij}$] is the total expected covariance between phenotypic values of species $i$ and $j$ (or giving the expected variance if $i = j$). The variances and covariances of and between species are given by the following (Felsenstein 1973):

$$\mathcal{S} = [s_{ij}] = \sum_{k=1}^{K} I_k \sigma_k^2 \qquad (A2)$$

In the above, $I_k$ is an indicator variable taking the value 1 if branch $k$ is part of the shared ancestry of species $i$ and $j$ in $\tau$ and taking the value 0 otherwise; $\sigma^2_k$ is the variance associated with the evolutionary process along the $k^{th}$ branch. The conditional probability density of the multivariate normal distribution is given by:

$$\mathcal{L}(\mathbf{y}|\mu, \mathcal{S}) = \frac{\exp\{-\frac{1}{2}[y-\mu]'\mathcal{S}^{-1}[y-\mu]\}}{\sqrt{(2\pi)^N |\mathcal{S}|}} \qquad (A3)$$

The total variance expected to accumulate along each branch $k$ ($\sigma^2_k$) is of key interest. In our Simpsonian model, we treat the total accumulated variance along each branch as the composite of a continuous diffusion and an instantaneous jump process. In the general model, we have

$$\sigma_k^2 = \tau_k \sigma_{W_k}^2 + n_k \sigma_{J_k}^2 \qquad (A4)$$

where $\tau_k$ is the length of the $k^{th}$ branch of $\tau$ (in units of time), $\sigma^2_{Wk}$ is the diffusion rate of a random-walk process associated with the branch, $n_k$ is the number of evolutionary jumps along the branch, and $\sigma^2_{Jk}$ is the variance associated with the jump process. This model describes a process whereby a trait experiences pulses of evolution in particular lineages (i.e., where $n_k$ is non-zero) superimposed on a random diffusion process. Landis et al. (2013) describe a similar model that differs in important ways: for instance, the Landis et al. (2013) model generates expected distributions of species' traits that have heavier tails than our multivariate normal.

The above describes the general model, where jumps may occur on any branch of the tree and where diffusion rates and jump variances vary across branches. We also consider many elaborations and restrictions of the general model. These simpler models are what are used in our data analysis (see *Material and Methods*). One such model, which we refer to as the free model, involves a tree-wide jump process (i.e., occurring at any branch in the tree) in combination with a global diffusion-rate parameter, $\sigma^2_W$, and a single pulse parameter, $\sigma^2_J$, the variance associated with each jump in the jump process. The simplest model is a global-rate random-walk process, where all branches share a rate parameter $\sigma^2_W$, and evolution along a single branch is modeled by $\sigma^2_k = \tau_k \sigma^2_W$ (i.e., all $n_k$ are zero; Cavalli-Sforza and Edwards 1967; Felsenstein 1973). Other model restrictions might entail splitting the tree a priori into distinct classes of branches, where each branch class has a unique diffusion-rate parameter (e.g., see O'Meara et al. 2006; Revell et al. 2012; Thomas and Freckleton 2012) and (or) jump process.

To efficiently compute the conditional model likelihood without the need to invert S -- see equation (A3) -- we follow FitzJohn's (2012) 'pruning' method (see also Felsenstein 1973 and Freckleton 2012). This pruning method requires a fully resolved tree. Under all the models considered here, the partial log-likelihood for the bifurcation that subtends species (or nodes) *i* and *j* is simply

$$\log_e \mathcal{L}_k(y_i, y_j | \sigma_i^2, \sigma_j^2) = -\frac{(y_i - y_j)^2}{2(\sigma_i^2 + \sigma_j^2)} - \frac{\log_e \{2\pi(\sigma_i^2 + \sigma_j^2)\}}{2} \quad (A5)$$

Each $\sigma^2$ term captures both the random-walk and jump process alone each branch, as described in equation (A4). This implementation is related to Felsenstein's (1985) method of contrasts insofar as phenotypic values at internal nodes are estimated using a phylogenetically weighted average of the values at subtended nodes (or tips):

$$E[y_k] = \frac{y_i \sigma_i^2 + y_j \sigma_j^2}{\sigma_i^2 + \sigma_j^2} \quad (A6)$$

Here, $k$ is the immediate ancestor of species $i$ and $j$. In order to account for estimation error associated with character values estimated for internal nodes, the expected variance along branch $k$ (i.e., the stem branch of descendants $i$ and $j$) must be scaled following Felsenstein (1985):

$$\sigma_k^{2'} = \sigma_k^2 + \frac{\sigma_i^2 \sigma_j^2}{\sigma_i^2 + \sigma_j^2} \quad (A7)$$

We conduct a postorder traversal to obtain the model likelihood $L$, summing the partial log-likelihoods computed for each internal node. Arriving at the root node $K$ (whose character value is a free parameter $\mu$), the partial log-likelihood is

$$\log_e \mathcal{L}_K(y_i, y_j | \sigma_i^2, \sigma_j^2, \mu) = -\frac{(E[y_K] - \mu)^2}{2\sigma_K^{2'}} - \frac{\log_e \{2\pi \sigma_K^{2'}\}}{2} \quad (A8)$$

assuming $\sigma^2{}_K = 0$. The quantity $\sigma^{2'}{}_K$ is therefore determined solely by the second term of equation (A4). The partial log-likelihood at the root is added to the sum of the partial log-likelihoods from equation (A5) for all *N*-2 bifurcations to obtain the model log-likelihood.

## APPENDIX B

**MCMC algorithm**

To fit this model to data, we treat the locations of evolutionary pulses as latent variables to be estimated by Markov chain Monte Carlo sampling. One can also imagine an approach that treats the arrival of evolutionary jumps as a parametric process (and see Landis et al. 2013 for just such an implementation). We use a zero-inflated log-uniform distribution for our prior probability mass on the total number of jumps across the tree. We apportion 1/3 of the total probability mass to the zero-jump class. In the accompanying software, this prior distribution on jumps (and all other prior distributions) can be flexibly defined. Following Simpson (1944), we assume quantum evolution to be a rare occurrence and we thus restrict our MCMC algorithm to sample no more than a single jump along any branch in the phylogeny at maximum. We use vague exponential prior densities for the diffusion and jump variances, both with mean $10^3$. These distributions are broad but appropriate for our log-transformed body shape data and with trees scaled to an arbitrary depth of 1 unit (see *Morphometric data and phylogeny*). We also use a vague prior for root state, $U(-\infty, \infty)$. The proposal density for moving and adding jumps is uniform across the collection of branches in the phylogeny and independent of branch lengths. For each proposal, we use a proposal width of 1. Other details of the MCMC algorithm follow Eastman et al. (2011). As shown in our empirical analyses, this method is easily extended to cases where there exists uncertainty in the phylogeny.

**Simulations**

We test the adequacy of our MCMC sampler to recover the generating process for datasets simulated under known conditions. We use two basic models under which to simulate: a single-rate Brownian-motion process and a jump-diffusion process. In all simulations, we have a single class of the diffusion parameter ($\sigma^2_W = 0.01$). We allow the jump variance ($\sigma^2_J$) to vary across simulations. For simplicity, we define α = $\sigma^2_J / \sigma^2_W$ (the ratio between the the jump variance and Brownian-motion variance) as a fundamental dynamic of the generating process. The value of α can be considered to be the relative time required under diffusive evolution to produce a comparable effect on phenotypic evolution as a single instantaneous pulse of evolution. We generate simulations under the conditions α ∈ (1, 10, 100, 1000). Using TREESIM (v1.5; Stadler 2011), APE (v3.0.1; Paradis et al. 2004), and GEIGER (v1.3-1; Harmon et al. 2008), we generate trees with two distinct shapes: a subset of trees were perfectly balanced with equal (unit) branch lengths; the remaining trees were generated under a birth-death process with speciation rate of 0.020 events per time unit and extinction rate of 0.015 events per time unit. To improve comparability of trees differing in shape, we transform the birth-death trees such that mean branch length is 1.0 time units (as in the perfectly balanced set of trees), while preserving the original shape of each rescaled tree. Number of tips for the set of balanced trees is 128 ($2^7$), 1024 ($2^{10}$), and 8192 ($2^{13}$); tree sizes for the birth-death trees are $10^2$, $10^3$, and $10^4$. For each tree type and size, we generate ten replicates. For datasets where α > 0, we simulate jumps such that the expected total number of evolutionary pulses is a given proportion of the total number of branches in the phylogeny ($f_J$): 0, 0.01, 0.025, 0.05, 0.10, or 0.20. Jumps are distributed across branches at random and with the restriction that no more than a single jump could be present along any given branch. We simulate a total of 6300 datasets: under the strict diffusion model ($f_J = 0$), we have 6 tree types, 10 tree replicates per tree, and 5 simulated datasets per tree; for jump-diffusion simulations ($f_J > 0$), we have 6 tree types, 10 tree replicates, 5 $f_J$, 4 α, and 5 simulated datasets per tree. We run the jump-diffusion MCMC sampler on each simulation for $2 \cdot 10^6$ generations, sampling every 500th generation. We retain the last half of samples for analysis, having allowed the chains to burn-in. The Heidelberger and Welch (1983) test, implemented in the R-package CODA (v0.14.7; Plummer et al. 2006), is used to confirm stationarity for a subset of randomly chosen Markov chains.

In general the MCMC sampler exhibits adequate performance in recovering the generating jump-diffusion process (e.g., Fig. A1). Performance of the MCMC sampler appears marginally better for more balanced trees (data not shown). We attribute this observation to the homogeneity of branch lengths in our balanced trees, as there is a slight tendency for the MCMC sampler to miss jumps occurring on longer branches. Along a single branch, the total expected variance of the trait evolutionary process is a combination of both the jump and diffusive processes ($\sigma^2_k = \tau_k \sigma^2_W + n_k \sigma^2_J$). All else being equal, a jump occurring along a short branch will be more easily detected: $\sigma^2_k$ becomes overwhelmed by the first term (i.e., $\tau_k \sigma^2_W$) on a long branch. We expect that especially where the variance of the jump process is similar to the total variance accumulated by the diffusive process ($\tau_k \sigma^2_W$), jumps may be missed. This is related to the observation of an underestimated total number of jumps for simulated datasets where α is small (Fig. A2). While simulations across both birth-death and balanced trees are pooled in Figure A2, the bias appears weaker for more balanced topologies. As jumps become frequent, the sampler frequently overestimates $\sigma^2_W$, especially where jump intensity (α) is moderate (Fig. A3). Especially for smaller trees and for simulations where the true jump variance is small, the median estimated jump variance across similar simulations is near 0 (Fig. A4), and these are many of the same cases under conditions where the number of jumps is underestimated (e.g., Fig. A2). In contrast, the sampler has a slight tendency to overestimate the known jump variance in larger trees, particularly where the true jump variance is moderate relative to the diffusion variance (e.g., 10 ≤ α ≤ 100; Fig. A4).

We use a branchwise measure to explore accuracy and precision of the MCMC sampler in localizing estimated jumps to particular branches of tree (Figs. A5-A7). This measure compares the posterior probability of an estimated jump at a particular branch to an 'accuracy threshold' to determine if the inference is accurate or inaccurate. Results show several desirable characteristics of the sampler. As the relative intensity of the jump process (α) increases, false negatives (i.e., incorrectly inferring the absence of a jump) quickly descend to low rates regardless of stringency of the accuracy threshold (Fig. A5). Rates of false positives (i.e., erroneously inferring the presence of a jump) are exceedingly low, revealing the sampler to be generally conservative (Fig. A6). Median false positive rates are often very well below 0.02 and are never higher than 0.05, even under the most liberal accuracy threshold (Fig. A6). As jumps become especially frequent, false-positive rates modestly increase, especially where the threshold for declaring a false positive is low (Fig. A6). Expectedly, the ability of the sampler to correctly localize jumps tends to increase with increasing rarity and intensity of the jump process and as the accuracy threshold becomes more lenient (Fig. A7).

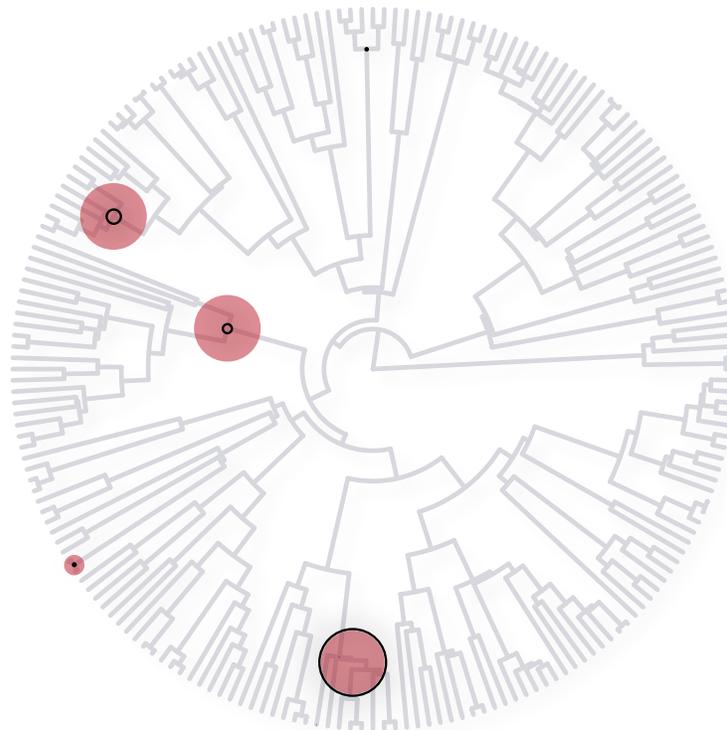

Figure A1. Illustration of algorithm performance from a single simulated tree and dataset. Simulated jumps are shown as black circles: circle size is proportional to the effect size of a true jump (i.e., the absolute value of the random Gaussian deviate drawn for each simulated jump). Reconstructed jumps are shown as red circles. Marginal posterior probabilities (proportional to size of shaded blue circles) of four estimated jumps range between 0.27 and 1.00, with the three most highly supported jumps 0.99 ≤ PP ≤ 1.00. Posterior probabilities of all other (unshown) estimated jumps are at or below 0.01. Jumps with the largest effect sizes are detected by the sampler, yet one other jump of minimal effect size (barely visible) is missed. This undetected jump had very little effect on the simulated process of phenotypic evolution and thus inclusion of a jump at this branch contributes little to the marginal likelihood of the model.

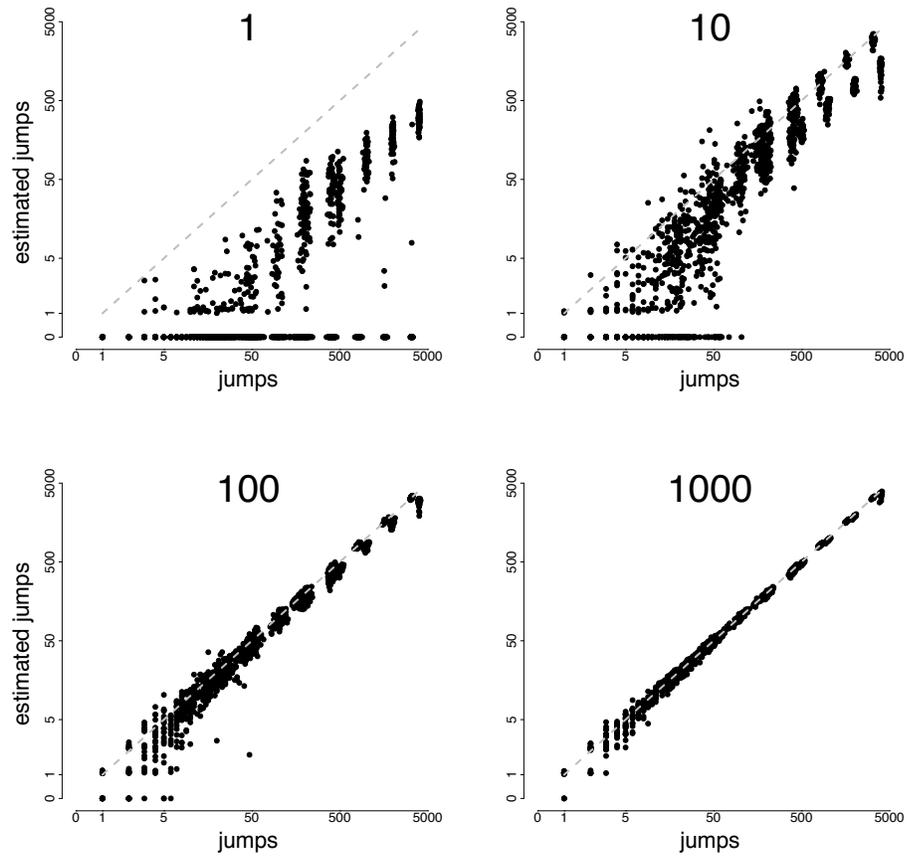

Figure A2. Plots showing the number of simulated jumps (x-axes) against the estimated number (y-axes). Panels show pooled sets of simulations: each panel includes estimates for replicate simulations under five jump rates ($f_J$) and six tree types. The relative importance of the jump process increases from leftmost to rightmost panels, with values of α shown at the top of each panel. The gray line -- present in each panel and most visible in the leftmost panel -- defines the expectation for estimated jumps for the sampler.

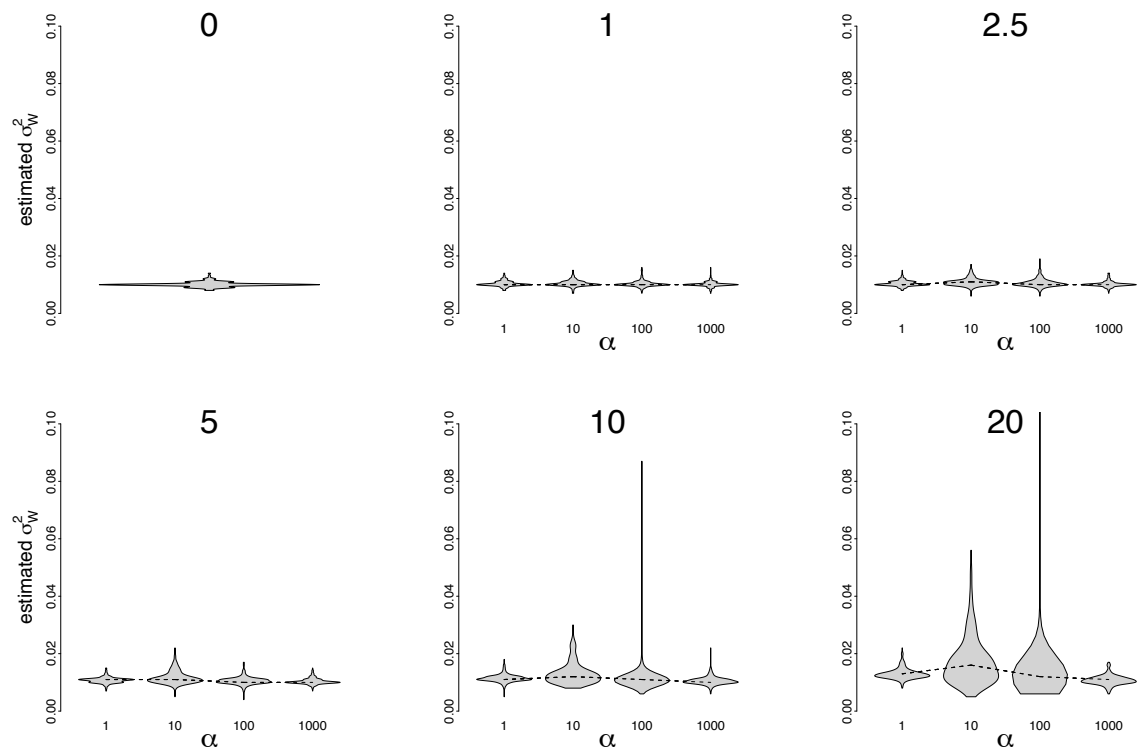

Figure A3. Estimates of the random-walk variance ($\sigma^2_W$) from simulated data. Violin plots depict the densities of estimates for each relative intensity of the jump process (i.e., $\alpha = \sigma^2_J / \sigma^2_W$), which increases along the x-axes. The simulated (true) random-walk variance is $\sigma^2_W = 0.01$. Values above each panel represent the stochastic proportion of branches in the tree for which a jump was simulated ($f_J$). The upper leftmost panel depicts estimates for simulations in the absence of jumps and thus $\alpha$ is zero. Median estimates of $\alpha$ are connected by dashed line in panels.

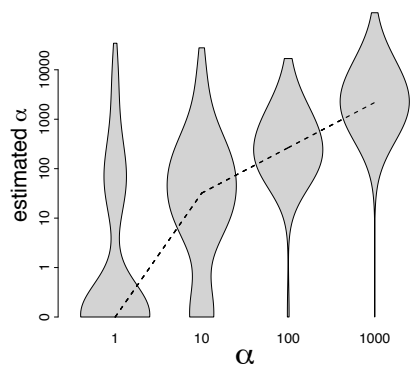

Figure A4. Estimates of the scaled jump variance ($\alpha$; y-axis) for four simulation conditions of increasing jump variance along the x-axis. All simulations use a diffusion variance of $\sigma^2_W = 0.01$. Results are shown only for the subset of simulations where jumps are included in the simulated process of evolution. Each violin plot pools results amongst several simulation conditions, spanning several tree types and jump frequencies, $f_J$ (see Fig. A2 caption).

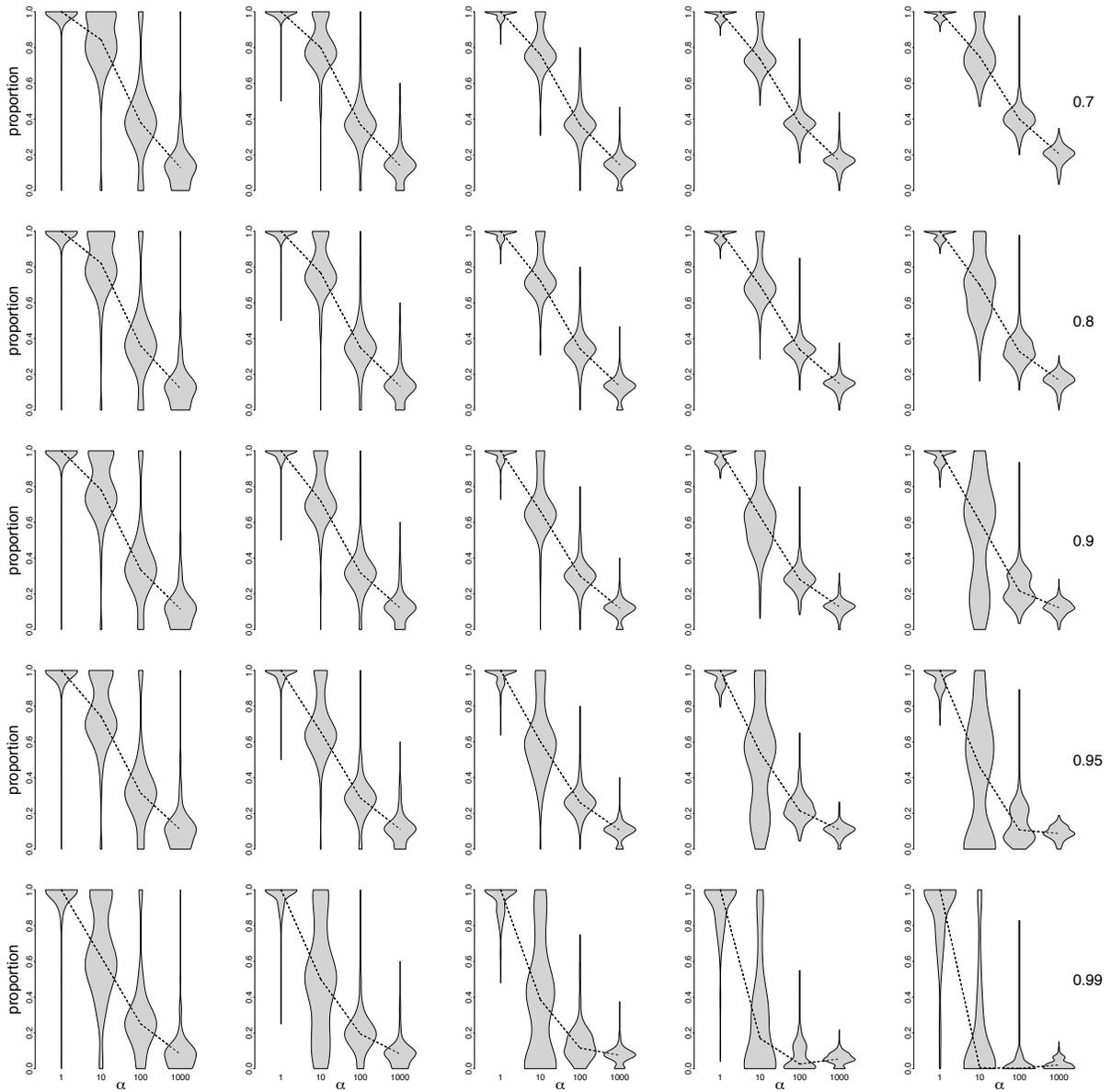

Figure A5. A measure of accuracy in estimating the jump process for simulated trees is plotted as a branchwise proportion of 'false negatives' (i.e., branches truly experiencing jumps but erroneously inferred to be without jumps). This measure is a treewide assessment of accuracy in estimating the true jump-diffusion process: estimates of jumps along each branch in the tree are compared to the known (simulated) process. Considering a branch that truly experienced a jump in simulation, an inference at a branch is deemed a false negative when the posterior probability for the absence of jumps along that branch exceeds a given threshold. We use a series of accuracy thresholds, given by values along the rightmost margin. Treewide accuracy is simply tabulated as the mean false-negative rate across all branches. Violin plots show the densities of treewide accuracies along the y-axes over the range of increasingly stringent thresholds and over a range of jump intensities (α, along the x-axes). Columns are increasing proportions of branches in simulated trees that have experienced an evolutionary pulse ($f_J$), from 0.01 (leftmost), 0.025, 0.05, 0.10, to 0.20 (rightmost). Results are depicted as the density of treewide false negatives for the subset of branches in simulated trees that truly experience a jump. Median rates of false negatives across jump intensities are connected by dashed line. y-axes range from 0 to 1.

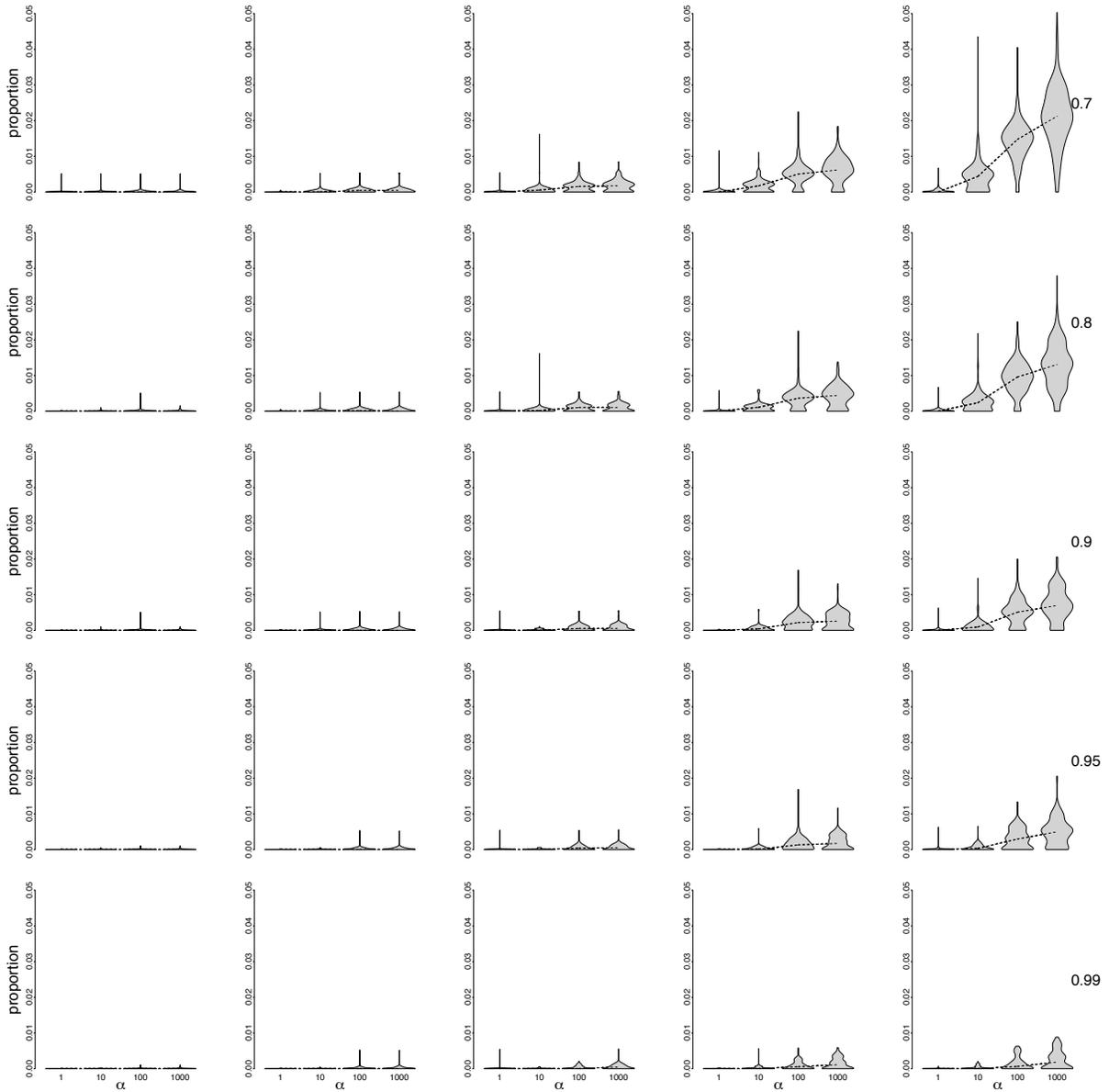

Figure A6. Panels summarize treewide proportions of 'false positives' (i.e., the frequency of branches lacking jumps in simulation but erroneously inferred to have experienced jumps). The subset of branches in each simulated tree that did not experience a jump are considered. Similar to Figure A5, y-axes represents densities of treewide proportions of false positives, using increasingly stringent thresholds for declaring an inferential error. Note that to clarify trends, y-axes range from a proportion of 0 to a maximum of 0.05. Other details follow Figure A5.

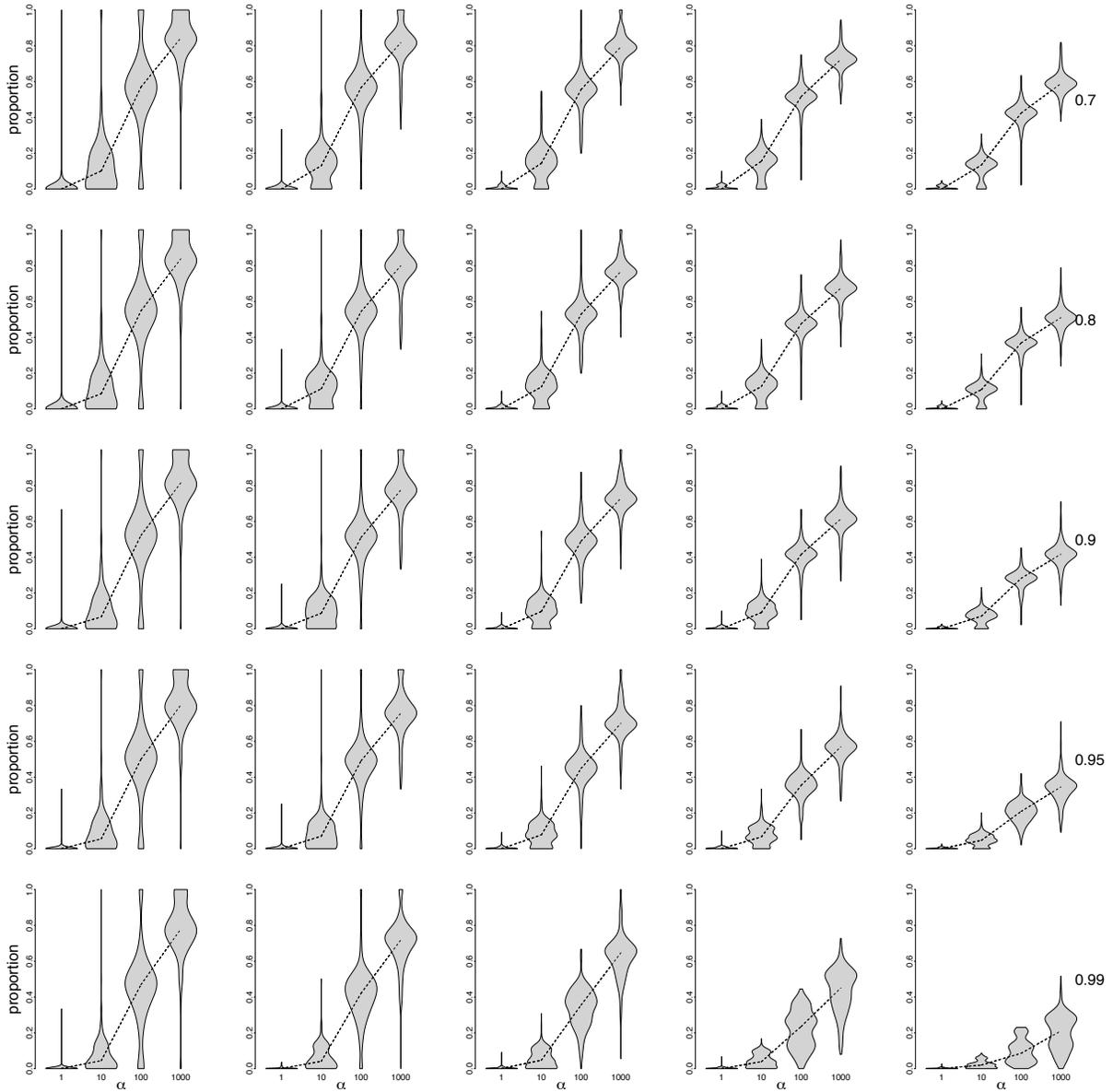

Figure A7. Panels summarize ability of the MCMC algorithm to accurately estimate the presence of jumps along branches truly experiencing jumps in simulated trees. Plotted results consider only the subset of branches that have truly experienced jumps in simulation. The y-axes represents treewide accuracy in recovering the simulated process. An inference at a branch is deemed accurate where the posterior probability of a jump exceeds a threshold (values indicated in the rightmost margin). Tolerances range from liberal (threshold = 0.70 in the uppermost panels) to stringent (threshold = 0.99 in the lowermost panels). Other details follow Figure A5.

**Simulations with the anole tree**

To investigate the ability of our sampler to recover known features of the jump-diffusion process on the *Anolis* tree, we generate a set of simulations that largely follow the procedures developed and described in the Simulations section. Rather than simulate trees, we instead use 100 random draws from the posterior density of trees for Greater Antillean anoles (see *Morphometric data and phylogeny*. We generate data under the same conditions described in *Simulations*. Patterns of sampler performance are similar between these two sets of simulations, and we highlight where algorithm performance is most discrepant. The proportion of branches that we identify as 'false positive' are very near to zero in the *Anolis* simulations as previously. Yet in general, the intensity of the jump process (α) needed to be greater in the *Anolis* simulations in order for proportions of 'false negatives' to be depressed and for 'true positives' to be elevated. This is largely attributed to the inability of the sampler to recover the proper number of jumps until α is quite large (Fig. A10). For many replicate simulations on the *Anolis* trees, α is severely underestimated (Fig. A11). Results for these simulations underscore the conservatism of the sampler and indicate that the jump process estimated for the empirical datasets in the Greater Antillean anoles may be underestimated in both rate and intensity.

Figure A8. Bivariate plot of species scores along the body size and first (PCI) principal-component axes. Convex hulls of the six ecomorph classes are plotted. Species labels are in the vicinity of points.

Figure A9. Bivariate plot of species scores along the body size and third (PCIII) principal-component axes. Details follow Figure A8.

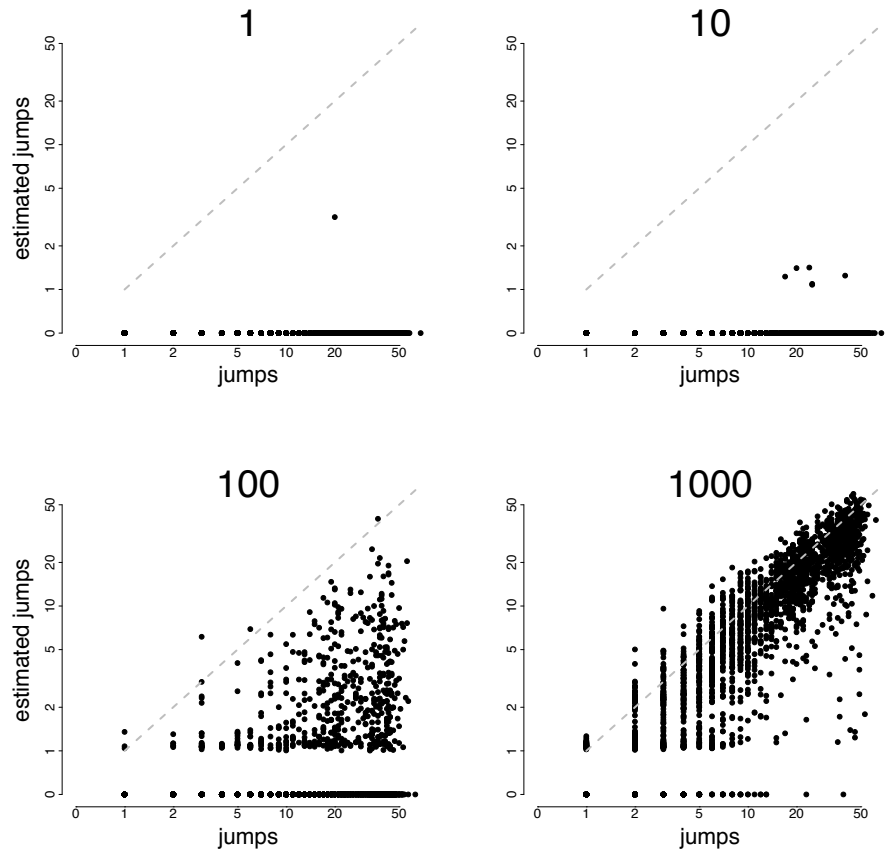

Figure A10. Plots depict the number of simulated jumps (x-axes) against the estimated number (y-axes), for simulations generated on the 100 taxon *Anolis* trees. Other details follow Figure A2.

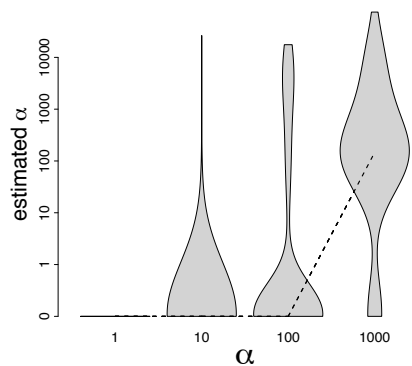

Figure A11. Plots depict variance of the simulated jump process ($\alpha = \sigma^2_J / \sigma^2_W$; x-axes) against estimates of the scaled jump variance (y-axes), for simulations generated on the 100 taxon *Anolis* trees. Other details follow Figure A4.